\pdfoutput=1
\documentclass[11pt,a4paper]{article}
\usepackage[margin=2.3cm]{geometry}
\usepackage{amsmath,amssymb}
\usepackage{graphicx}
\usepackage{booktabs}
\usepackage{placeins}
\usepackage{cite}
\usepackage[colorlinks=true,citecolor=blue,linkcolor=blue,urlcolor=blue]{hyperref}
\graphicspath{{figures/}}
\allowdisplaybreaks

\newcommand{\Om}{\Omega_{m0}}
\newcommand{\Ok}{\Omega_{k0}}
\newcommand{\Od}{\Omega_{d}}
\newcommand{\wde}{w_d}

\title{When the future decides the present: a consistency test for new Tsallis holographic dark energy}

\author{P. Suresh Kumar$^{1}$, Bramha Dutta Pandey$^{1}$, Umesh Kumar Sharma$^{2,}$\thanks{Corresponding author: sharma.umesh@gla.ac.in} Pankaj$^{3}$\\[3mm]
\small $^{1}$Department of Supportive Requirements, University of Technology and Applied\\
\small Sciences-Salalah, Sultanate of Oman\\
\small $^{2}$ Department of Mathematics, Institute of Applied Sciences and Humanities, GLA University, Mathura,\\
\small Uttar Pradesh 281406, India\\
\small $^{3}$School of Sciences, IILM University, Knowledge Park-II, Greater Noida 201310, India,\\
\\[2mm]
\small E-mails: sureshharinirahav@gmail.com, bdpandey05@gmail.com,\\
\small sharma.umesh@gla.ac.in, pankaj.fellow@yahoo.co.in,}
\date{}

\begin{document}
\maketitle

\begin{abstract}
New Tsallis holographic dark energy (NTHDE) is obtained from the Tsallis entropy of the
cosmological horizon. In several studies, its exponential entropy is replaced by a three-term
series, and the future event horizon is used as the infrared cutoff. We test both choices
with DESI DR2 baryon acoustic oscillations, Pantheon+SH0ES supernovae and cosmic
chronometers, in flat and curved universes and without cosmic microwave background data. We
show that the future-event-horizon cutoff has a self-consistency condition that is usually not
checked: the horizon length obtained from the present dark-energy density must equal the event
horizon of the expansion predicted by the model. This condition always holds in
standard holographic dark energy (HDE). In NTHDE, it fails for about half of the parameter
region allowed by the data, where the future dark-energy density grows very rapidly. A
fixed-point analysis of the late-time dynamics explains this behaviour. With the
condition imposed, NTHDE gives $H_0=72.2\pm0.9$ km s$^{-1}$ Mpc$^{-1}$, a spatial curvature
consistent with zero and a present equation of state of $-0.97\pm0.05$. The full and truncated
entropies fit the data equally well, but the truncation is not a controlled approximation, and
the Bayesian evidence does not prefer either form over $\Lambda$CDM. We get the same results
with the Union3 and DES-Dovekie supernova samples. With the apparent horizon as cutoff, NTHDE
behaves like matter at high redshift and is strongly disfavoured. This test can be applied to
any HDE model with a future-event-horizon cutoff.
\end{abstract}


\section{Introduction}
\label{sec:intro}

The accelerated expansion of the Universe at late times
\cite{SupernovaSearchTeam:1998fmf,SupernovaCosmologyProject:1998vns} is still not well
understood. The cosmological constant is the simplest explanation, but it has the fine-tuning
and coincidence problems \cite{Weinberg:1988cp,Padmanabhan:2002ji}. Also, the values of the
Hubble constant obtained from early-Universe and late-Universe data do not agree
\cite{DiValentino:2021izs,Riess:2020fzl}. For these reasons, dynamical dark-energy models are
still widely studied. The latest baryon acoustic oscillation (BAO) data from DESI, combined
with supernovae and the cosmic microwave background (CMB), prefer evolving dark energy at the
$2$--$4\sigma$ level. The strength of this preference depends on the supernova sample used
\cite{DESI:2025zgx,Wang:2026dde}.

Holographic dark energy (HDE) is based on the holographic principle
\cite{Susskind:1994vu,Bousso:2002ju}. According to this principle, the vacuum energy in a
region of size $L$ cannot be larger than the energy of a black hole of the same size. This
gives $\rho_d=3c^2M_p^2L^{-2}$ \cite{Li:2004rb}. If the Hubble radius is used as the cutoff,
the model does not give accelerated expansion \cite{Hsu:2004ri}. With the future event
horizon, it does \cite{Li:2004rb}. However, the dark-energy density then depends on the future
expansion of the Universe. This causality problem has led to other cutoffs, such as the age of
the Universe \cite{Cai:2007us} and the Ricci scale \cite{Gao:2007ep}. In addition, gravitating
systems have long-range correlations, so the horizon entropy need not follow the
Bekenstein--Hawking area law \cite{Bekenstein:1973ur}. Generalised entropies therefore give a
whole family of holographic models \cite{Tavayef:2018xwx}. The thermodynamic viability of
these models is also being studied \cite{Cruz:2026gsl}.

New Tsallis holographic dark energy (NTHDE) was proposed by Moradpour \emph{et al.}
\cite{Moradpour:2020dfm}. They applied Tsallis statistics with equally probable horizon states
to the Bekenstein entropy. This gives an exponential entropy and an exponential dark-energy
density, with the apparent horizon as the cutoff. In later works, the entropy was expanded up
to second order in its deformation parameter. The resulting three-term density was studied with
the event horizon and the apparent horizon, in flat and curved universes
\cite{Pandey:2021fvr,Pankaj:2022lnk,Pankaj:2021nkg,Nisha:2022mpl}. Two questions are still
open. The first is whether the truncated density is a good approximation for the parameter
values allowed by the data, because the expansion parameter need not be small there. The
second is which infrared cutoff is preferred by the data. This is important because curvature
and an evolving equation of state change distances in a similar way
\cite{Huang:2004ai,Setare:2006wh}, and non-flat holographic models fitted to DESI DR2 can
prefer a non-zero curvature \cite{Shen:2026ihde}.

We have used the same data and analysis method in two earlier works: for NTHDE with the
Granda--Oliveros cutoff in a curved universe \cite{Sharma:2026goNTHDE}, and for Kaniadakis
holographic dark energy with the future event horizon \cite{Kumar:2026khde}. The present paper
is different in two ways. It uses the future event horizon for NTHDE, and it studies the
self-consistency of this cutoff. The Granda--Oliveros cutoff depends only on $H$ and $\dot H$,
so it has no self-consistency problem of this kind.

In this work, we study both questions with the same data and likelihoods for all models. We
use Pantheon+SH0ES supernovae \cite{Scolnic:2021amr,Brout:2022vxf,Riess:2021jrx}, cosmic
chronometers (CC), which measure $H(z)$ without a distance ladder
\cite{Jimenez:2001gg,Moresco:2012jh,Moresco:2016mzx,Moresco:2022phi}, and DESI DR2 BAO data
\cite{DESI:2025zgx}. The sound horizon $r_d$ and the absolute magnitude of the supernovae are
kept free, so no information calibrated by the CMB is used. We compare eleven models: the full
and the truncated NTHDE densities with the event horizon and the apparent horizon, in flat and
curved universes, together with HDE and flat and curved $\Lambda$CDM. For this, we use maximum
likelihood, Markov-chain Monte Carlo (MCMC) posteriors, information criteria and the Bayesian
evidence. We also show that the future-event-horizon cutoff needs a self-consistency
condition, and that this condition restricts the NTHDE parameter space. We check our
conclusions with the Union3 and DES-Dovekie supernova samples, with a correlated systematic
error in the chronometer data, and with a physically motivated prior on the matter density.
Finally, we reconstruct $H(z)$ with neural networks, without assuming a cosmological model
\cite{Seikel:2012uu,Mukherjee:2024ryz,Maqsood:2026thermo,Favale:2023gqk} (Appendix~\ref{app:ann}).

The paper is organised as follows. In Sect.~\ref{sec:model}, we present the full and truncated
densities, the two cutoffs and the self-consistency condition. Section~\ref{sec:data}
describes the data and the statistical methods. The results are given in
Sect.~\ref{sec:results}, and they are discussed in Sect.~\ref{sec:discussion}.
Section~\ref{sec:conclusions} gives our conclusions.

\section{New Tsallis holographic dark energy}
\label{sec:model}

\subsection{Full and truncated densities}
\label{sec:densities}

When all microstates are equally probable, the Tsallis entropy of a horizon with Bekenstein
entropy $S_{BH}$ is \cite{Moradpour:2020dfm}
\begin{equation}
S_T=\frac{1}{\delta_T}\left[\exp\left(\delta_T S_{BH}\right)-1\right],
\label{eq:ST}
\end{equation}
where $\delta_T=1-Q$ is the non-extensivity parameter, and $S_T\to S_{BH}$ as $\delta_T\to0$.
For a cutoff $L$, we have $S_{BH}=\pi L^2/G=8\pi^2M_p^2L^2$. Using the holographic relation
$\rho_d\propto S/L^4$, with a normalisation that gives HDE for $\delta_T\to0$, we get the full
NTHDE density
\begin{equation}
\rho_d=\frac{3c^2M_p^2}{L^2}\,\frac{e^{u}-1}{u},\qquad u\equiv\delta_T S_{BH}=\beta L^2,
\label{eq:full}
\end{equation}
where $L$ is measured in units of $H_0^{-1}$, and $\beta\equiv8\pi^2\delta_TM_p^2H_0^{-2}$ is the
dimensionless deformation parameter. The series $(e^u-1)/u=1+u/2+u^2/6+O(u^3)$ gives three
terms: $3c^2M_p^2/L^2$, a constant, and a term that grows as $L^2$. If their coefficients are
written with a single parameter $\delta$, we get the truncated density used in
Refs.~\cite{Pandey:2021fvr,Pankaj:2022lnk,Pankaj:2021nkg,Nisha:2022mpl},
\begin{equation}
\rho_d=\frac{3c^2M_p^2}{L^2}+\delta M_p^4+\delta^2M_p^6L^2 .
\label{eq:trunc}
\end{equation}
We note two points. First, the coefficients of the constant term and the $L^2$ term in the
series are $3c^2\beta/2$ and $c^2\beta^2/2$ (in units $M_p=H_0=1$). Writing them as $\delta$
and $\delta^2$ is correct only when $c^2=2/9$. Hence, Eq.~(\ref{eq:trunc}) is a separate
three-term model and not a controlled truncation of Eq.~(\ref{eq:full}). Second, the series is
accurate only for $u\ll1$, but the data allow $u=O(1)$ (Sect.~\ref{sec:fullvstrunc}).
Therefore, we take Eq.~(\ref{eq:full}) as the NTHDE model and Eq.~(\ref{eq:trunc}) as the
published truncated form, and we fit both. Both reduce to HDE when the deformation goes to
zero. The Bekenstein entropy of a cosmological horizon is of order $10^{122}$, so $u=O(1)$
means $\delta_T\sim10^{-122}$ in Planck units. Thus, the model replaces the small value of the
cosmological constant by a small value of the non-extensivity parameter. This fine-tuning
problem is common to other generalised-entropy models, and we do not try to solve it here.

\subsection{Non-flat background and infrared cutoffs}
\label{sec:cutoffs}

For a Friedmann--Robertson--Walker universe with spatial curvature $k$, containing dark energy
and pressureless matter, the Friedmann equation is
\begin{equation}
3M_p^2H^2+3M_p^2\frac{k}{a^2}=\rho_m+\rho_d,\qquad \Od+\Omega_m=1+\Omega_k ,
\label{eq:fried}
\end{equation}
where $\Omega_i=\rho_i/(3M_p^2H^2)$. As in earlier NTHDE papers, we use $\Omega_k=+k/(a^2H^2)$,
so that $\Ok>0$ means a closed universe. Matter is conserved, $\rho_m\propto a^{-3}$. Dark
energy is also conserved, and its equation of state parameter and the deceleration parameter
are
\begin{equation}
\wde=-1-\frac13\frac{\mathrm{d}\ln\rho_d}{\mathrm{d}x},\qquad
q=-1-\frac{\mathrm{d}\ln H}{\mathrm{d}x},\qquad x=\ln a .
\label{eq:wq}
\end{equation}
We consider two cutoffs.

\emph{Future event horizon (EH).} In this paper, ``event horizon'' and EH always mean the
future event horizon. It is given by $L=a\,r(t)$, with $r=\sin y/\sqrt k$
($\sinh y/\sqrt{|k|}$ for $k<0$) and $y=\sqrt{|k|}\int_t^\infty\mathrm{d}t'/a$
\cite{Li:2004rb,Huang:2004ai}. Taking the time derivative, we get $\dot L=HL-\cos y$
($\cosh y$ for $k<0$), that is,
\begin{equation}
\frac{\mathrm{d}L}{\mathrm{d}x}=L-\frac{\cos y}{H},\qquad
\cos y=\sqrt{1-\frac{kL^2}{a^2}} .
\label{eq:EH}
\end{equation}
For a given value of $\Od$ today, the relation $\rho_d(L_0)=3M_p^2H_0^2\Od(0)$ gives $L_0$. We
take the root that reduces to HDE. Then Eqs.~(\ref{eq:fried}) and (\ref{eq:EH}) are
integrated towards the past.

\emph{Apparent horizon (AH).} Here $L=R_A=(H^2+k/a^2)^{-1/2}$. This cutoff was used in the
original NTHDE paper \cite{Moradpour:2020dfm} and in Ref.~\cite{Pankaj:2021nkg}. In this case,
the Friedmann equation becomes an algebraic equation in $L$, $3M_p^2/L^2=\rho_m+\rho_d(L)$,
which has only one root at each redshift. The closure condition today then fixes $c$, so $c$
is a derived parameter.

For both cutoffs, we solve the equations directly for $L$ instead of $\Od$. This method works
for any $\rho_d(L)$, and we do not need to solve the density relation for $L$ at each step.
For the truncated model, it agrees with the $\Od$-based method to $2\times10^{-6}$
(Sect.~\ref{sec:validation}).

\subsection{Self-consistency of the future-event-horizon cutoff}
\label{sec:consistency}

The event horizon depends on the whole future of the model. In practice, however, the
evolution is started from the value $L_0$ obtained from today's dark-energy density. These two
must agree. In the flat case, Eq.~(\ref{eq:EH}) gives $\mathrm{d}(L/a)/\mathrm{d}x=-1/(aH)$,
so that
\begin{equation}
\frac{L}{a}=\frac{L_0}{a_0}-\chi(x),\qquad \chi(x)=\int_0^x\frac{\mathrm{d}x'}{aH},
\label{eq:La}
\end{equation}
and the solution is the true event horizon of the model only if $L/a\to0$, that is,
$L_0/a_0=\chi_\infty$. With curvature, the condition becomes
$L_0/a_0=\sin(\sqrt k\chi_\infty)/\sqrt k$ (or the $\sinh$ form). We measure it by the
mismatch
\begin{equation}
\mu\equiv\frac{L_0}{L_{\rm EH}}-1,\qquad
L_{\rm EH}=a_0\,\frac{\sin\left(\sqrt{k}\,\chi_\infty\right)}{\sqrt{k}},
\label{eq:mu}
\end{equation}
which we compute by solving the model forward in time. For HDE, $\mu=0$ always, because the
late-time attractor gives $L/a\to0$ for any initial condition. For NTHDE, this is not
guaranteed. For any holographic density of the form $\rho_d\propto S(A)/A^2$, we have
$\mathrm{d}\rho_d/\mathrm{d}L\propto k_{\rm eff}-2$, where
$k_{\rm eff}\equiv\mathrm{d}\ln S/\mathrm{d}\ln A$ is the effective entropy exponent. For the
full density, $k_{\rm eff}=ue^u/(e^u-1)$. It increases from $1$ and becomes $2$ at
$u_\ast=1.594$. For larger $u$, $\rho_d$ increases with $L$. If $u$ crosses $u_\ast$ in the
future while $L$ is still growing, the dark energy becomes strongly phantom, $L/a$ stops at a
non-zero value, and the solution is no longer an event horizon.

This behaviour is easy to understand at late times, when matter and curvature can be
neglected and $H^2\simeq\rho_d/(3M_p^2)$. Then Eq.~(\ref{eq:EH}) becomes
$\mathrm{d}\ln L/\mathrm{d}x=1-1/(c\sqrt{g(u)})$, with $g(u)=(e^u-1)/u$. For HDE, $g=1$, and
$L/a\propto a^{-1/c}\to0$ for every initial condition. This is why $\mu=0$ always for HDE. For
$\beta>0$ and $c<1$, there is a de Sitter fixed point at $c^2g(u_{\rm dS})=1$, where the
horizon is equal to the Hubble radius. This fixed point is a repeller. Solutions that stay
below it go to the HDE attractor and are self-consistent. Solutions that cross it are pushed to
$\mathrm{d}\ln L/\mathrm{d}x\to1$, so that $L/a$ tends to a constant. For $c\ge1$, there is no
fixed point and every solution finally behaves in this way, but the mismatch is exponentially
small when $c$ is close to $1$. This explains the threshold in Fig.~\ref{fig:consistency}
(left): $\mu$ is exactly zero below the separatrix and grows continuously above it. The simple
rule ``$c<1$ and the future solution stays below $u_{\rm dS}$'' agrees with the full numerical
test for $91\%$ of the posterior samples. The other samples lie close to the separatrix, whose
position is shifted by the matter era and by curvature. For this reason, we use the full
numerical test. The forward integration is done up to $a=e^{12}\simeq1.6\times10^5$. When we
extend it to $a=e^{20}$, the result does not change for any of the $1500$ test samples. Like
the cutoff itself, the test assumes that the model is valid for the whole future evolution. We
require $|\mu|<10^{-3}$. This tolerance is set by the numerical accuracy of the forward
integration. Parameter values that do not satisfy this condition are treated as outside the
model, in the same way as values for which $L_0$ does not exist.

The relation $\rho_d(L_0)=3M_p^2H_0^2\Od(0)$ has a second root, on the branch where $\rho_d$
increases with $L$ ($u_0>u_\ast$). Since $g(u_\ast)=2.46$, this root lies above the repeller
when $c>0.64$, which is true for $94\%$ of the posterior. In that case, it never gives a
self-consistent solution. For smaller $c$ it may do so, but these solutions do not reduce to
HDE when $\beta\to0$, so we do not consider them. We will also use the following result from
Eq.~(\ref{eq:wq}). Along any solution, $\rho_d'=(\mathrm{d}\rho_d/\mathrm{d}L)L'$, so
$\wde=-1$ exactly where $k_{\rm eff}=2$ (for $L'\neq0$).

\section{Data and methodology}
\label{sec:data}

\subsection{Data}

\emph{Supernovae.} We use the Pantheon+SH0ES sample
\cite{Scolnic:2021amr,Brout:2022vxf,Riess:2021jrx}. It contains $1657$ supernovae with
$z_{\rm HD}>0.01$ or in Cepheid host galaxies ($77$ calibrators), with the full statistical and
systematic covariance matrix $\mathbf C$. The residuals are $\Delta_i=\mu_i-\mu_i^{\rm Ceph}$
for the calibrators and $\mu_i-\mu_{\rm model}(z_i)$ for the other supernovae. A common
magnitude offset is marginalised analytically,
$\chi^2_{\rm SN}=\Delta^T\mathbf C^{-1}\Delta-(\mathbf 1^T\mathbf C^{-1}\Delta)^2/(\mathbf 1^T\mathbf C^{-1}\mathbf 1)$.
In this way, the calibration uncertainty is included in $H_0$.

\emph{Cosmic chronometers.} We use $32$ differential-age measurements of $H(z)$ in the range
$0.07\le z\le1.965$
\cite{Simon:2004tf,Stern:2009ep,Zhang:2012mp,Moresco:2012jh,Moresco:2015cya,Moresco:2016mzx,Ratsimbazafy:2017vga,Borghi:2021rft},
with a diagonal covariance. We do not use the radial-BAO value $H(0.57)$, to avoid counting the
same information twice with DESI.

\emph{BAO.} We use the $13$ DESI DR2 measurements in the range $0.295\le z\le2.33$ (six
$D_M/r_d$, six $D_H/r_d$ and one $D_V/r_d$), with their full covariance \cite{DESI:2025zgx}.
The sound horizon is kept free. If we used the Planck value \cite{Planck:2018vyg}, our test of
non-standard dark energy at late times would depend on $\Lambda$CDM physics in the early
Universe.

We checked that each likelihood gives the published result for flat $\Lambda$CDM. The
supernovae give $\Omega_m=0.3308\pm0.0180$ and $H_0=73.53\pm1.02$ (published values:
$0.334\pm0.018$ and $73.04\pm1.04$ \cite{Brout:2022vxf,Riess:2021jrx}). DESI DR2 gives
$\Omega_m=0.2975$, the same as the published value, and the chronometers give
$\Omega_m=0.320\pm0.058$ and $H_0=68.1\pm3.0$ \cite{Moresco:2016mzx}. In this paper, $H_0$ is
always in units of km s$^{-1}$ Mpc$^{-1}$.

\subsection{Models, priors and sampling}

The eleven models and their parameters are as follows: flat $\Lambda$CDM $(\Om,H_0,r_d)$;
curved $\Lambda$CDM (with $\Ok$ added); curved HDE with the event horizon
$(c,\Om,\Ok,H_0,r_d)$; full and truncated NTHDE with the event horizon, flat and curved (with
$\beta$ or $\delta$ added); and full and truncated NTHDE with the apparent horizon, flat and
curved $(\beta\ \text{or}\ \delta,\Om,[\Ok],H_0,r_d)$, where $c$ is derived. We use flat
priors: $c\in[0.3,2]$, $\beta\in[0,5]$ and $\delta\in[0,1.2]$ (event horizon),
$\beta\in[0,20]$ and $\delta\in[0,5]$ (apparent horizon), $\Om\in[0.1,0.5]$,
$\Ok\in[-0.3,0.3]$, $H_0\in[60,85]$ and $r_d\in[120,180]$ Mpc. Only parameter values for which
the model has a real solution are allowed.

The posteriors are sampled with \texttt{emcee} \cite{Foreman-Mackey:2012any}, using $48$
walkers and $2\times10^4$ steps per chain. The first quarter of each chain is removed as
burn-in. A chain is extended automatically until its length is more than $50$ autocorrelation
times $\tau$ for every parameter. All joint chains converged in the first run. For every
parameter, the chain length is more than $60\tau$ (the slowest is $c$ in the full NTHDE
chains, and most parameters are above $100\tau$), and the Gelman--Rubin statistics are
$\hat R\le1.004$. The supernova-only chain of the main model was extended to
$2.65\times10^4$ steps ($55\tau$). After $6\times10^4$ steps, the CC+BAO-only chain reaches
$43\tau$ for $\beta$, which these data hardly constrain, and more than $59\tau$ for the other
parameters, with $\hat R\le1.001$. For the event-horizon NTHDE models, we impose the
self-consistency condition of Sect.~\ref{sec:consistency} as follows. We compute $\mu$ for
$1.2\times10^4$ posterior samples of each model and keep the samples with $|\mu|<10^{-3}$.
Since the condition is a strict limit on the model domain, the kept samples are exact samples
of the restricted posterior. The maximum-likelihood points are found by optimisation from many
starting points, with the condition imposed where it applies.

\subsection{Model comparison}

We compare each model with flat $\Lambda$CDM using
$\Delta\mathrm{AIC}=\Delta\chi^2_{\min}+2\Delta k$ \cite{Akaike:1974vps} and
$\Delta\mathrm{BIC}=\Delta\chi^2_{\min}+\Delta k\ln N$ \cite{Schwarz:1978tpv,Liddle:2007fy},
with $N=1702$ data points. The supernova magnitude is a nuisance parameter common to all
models. We also give the deviance information criterion \cite{Spiegelhalter:2002yvw},
$\mathrm{DIC}=\bar D+p_V$ with $p_V=\mathrm{var}(D)/2$ \cite{Gelman:2013bda}. We use this
variance form of the effective number of parameters because several posteriors are cut by
prior or validity limits, and the original form $\bar D-D(\bar\theta)$ is not reliable in such
cases.

The Bayesian evidence $Z$ is computed by nested sampling \cite{Skilling:2006gxv} with
\texttt{dynesty} \cite{Speagle:2019ivv} ($500$ live points, stopping at $\Delta\ln Z=0.1$).
This gives $\ln Z$ with an accuracy of about $\pm0.2$. For the nearly Gaussian posteriors
($\Lambda$CDM, curved $\Lambda$CDM and HDE), the result agrees with the Laplace approximation
within $0.2$. For the event-horizon NTHDE models, the evidence of the self-consistent model is
$Z_c=Z\,f_c$, where $f_c$ is the fraction of the posterior that satisfies the condition. In
this method, the prior is uniform over the whole parameter box, and parameter values with no
solution or with no self-consistent solution have zero likelihood. So the evidence includes a
penalty for this lost volume. Another choice is to normalise the prior only over the allowed
region. This increases $\ln Z$ by $-\ln V_a$, where $V_a$ is the allowed fraction of the prior
box. $V_a$ depends on the priors but not on the data. By Monte Carlo sampling of the prior, we
find $V_a=0.12$ for the full and $0.21$ for the truncated event-horizon NTHDE models, $0.96$
for HDE, $0.21$ for the truncated apparent-horizon models, and $1$ for $\Lambda$CDM and the
full apparent-horizon models. We give $\ln B$ for both choices.

We interpret $\ln B=\ln Z-\ln Z_{\Lambda\rm CDM}$ using the Jeffreys scale: $|\ln B|<1$ is
inconclusive, $1$--$2.5$ is weak, $2.5$--$5$ is moderate and $>5$ is strong
\cite{Kass:1995loi,Trotta:2008qt}. Since most of the $N$ data points are supernovae, the BIC
penalty is large, so we give more importance to the evidence. To test how $\ln B$ depends on
the prior range of the deformation parameter, we use the posterior samples and halve and
double this range. Finally, we measure the agreement between the supernova data and the CC+BAO
data by the loss in goodness of fit when they are combined,
$\Delta Q=\chi^2_{\min,\rm joint}-\chi^2_{\min,\rm SN}-\chi^2_{\min,\rm CC+BAO}$. We compare
$\Delta Q$ with a $\chi^2$ distribution whose number of degrees of freedom is equal to the
number of parameters constrained by both data sets \cite{Raveri:2018wln}.

\subsection{Additional tests}
\label{sec:robust_method}

We repeat the model comparison for four changes in the analysis. (i) and (ii): Pantheon+SH0ES
is replaced by Union3 \cite{Rubin:2023ovl} ($22$ binned distances with full covariance) or by
DES-Dovekie \cite{Popovic:2025dov} ($1820$ supernovae with the inverse total covariance).
These samples have no Cepheid calibration. Therefore, the absolute magnitude is marginalised
as a free offset, and $H_0$ is fixed by the chronometers and BAO. For flat $\Lambda$CDM and
supernovae alone, our Union3 likelihood gives $\Omega_m=0.356^{+0.028}_{-0.026}$, the same as
the published value \cite{Rubin:2023ovl}. Our DES-Dovekie likelihood gives $0.329\pm0.015$
(published value: $0.330\pm0.015$ \cite{Popovic:2025dov}). (iii) A fully correlated $5\%$
systematic error is added to the chronometer covariance. This is a simple way to include the
stellar-population systematics of Ref.~\cite{Moresco:2020fbm}. (iv) For the apparent-horizon
models, the lower limit of the matter-density prior is reduced to $\Om\ge0.04$, which is the
baryon density from Big Bang nucleosynthesis. In each case, we use maximum likelihood from many
starting points and nested sampling with the same settings as above. The posteriors are taken
from the nested-sampling samples, and the self-consistency condition is imposed on $4000$
posterior samples of each event-horizon NTHDE model. The information criteria use $N=67$
(Union3) and $N=1865$ (DES-Dovekie).

\subsection{Validation of the code}
\label{sec:validation}

Before every fit, the code passes nine checks. The event-horizon solutions agree with the
$\Od$-based evolution of the truncated model to $2\times10^{-6}$ in $H(z)$. The full density
goes to HDE when $\beta\to0$ ($3\times10^{-10}$) and agrees with its series expansion. Energy
conservation is satisfied: the equation of state from Eq.~(\ref{eq:wq}) agrees with
$\rho_d'/\rho_d$ from Eq.~(\ref{eq:EH}) to $6\times10^{-8}$. The apparent-horizon solutions
satisfy the Friedmann equation to $3\times10^{-14}$. The distances agree with direct numerical
integration to $10^{-6}$. The likelihoods give the same best fits as our earlier code
($\chi^2=1485.43$ for the truncated model and $1487.61$ for $\Lambda$CDM). The integration
settings used in the sampling change the distance moduli by at most $4\times10^{-6}$ mag.

\emph{Use of AI tools.} The authors used the AI assistant Claude (Anthropic) to help with code
development, numerical checks, and drafting and editing of the text. The authors checked all
results and the full text, and they take full responsibility for the content of this article.

\section{Results}
\label{sec:results}

\subsection{Effect of the self-consistency condition}

\begin{figure*}[t]
\centering
\includegraphics[width=\textwidth]{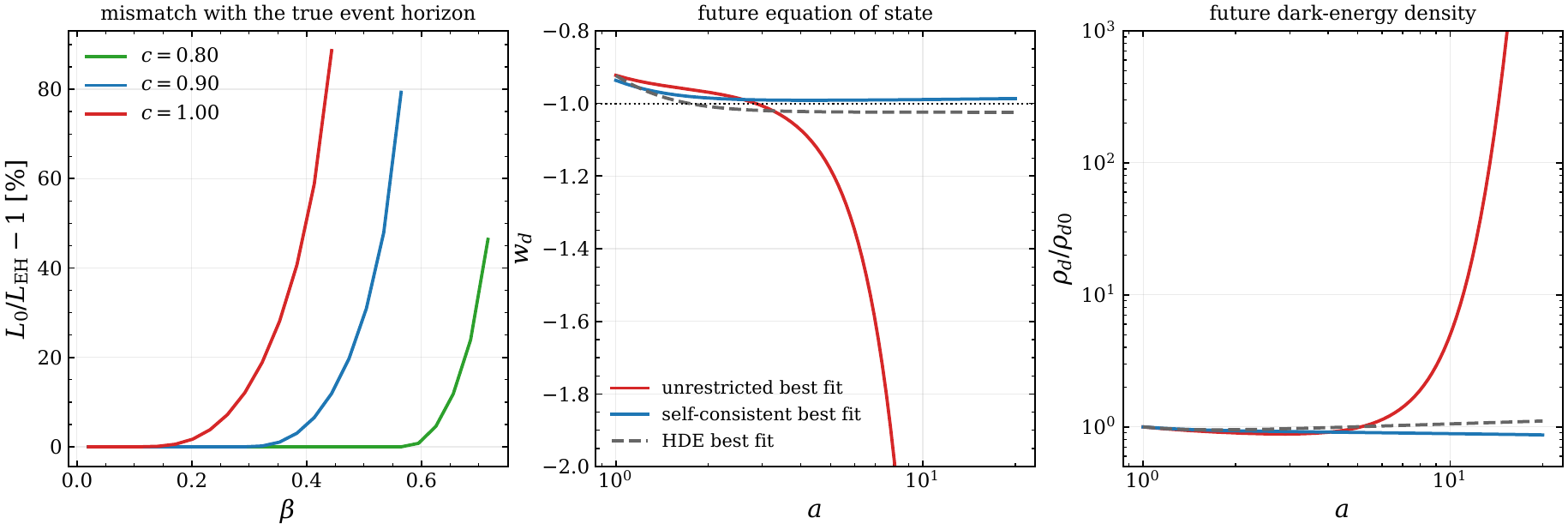}
\caption{Self-consistency of the future-event-horizon cutoff for the full NTHDE model. Left:
mismatch $\mu$ between $L_0$ and the true future event horizon of the model,
Eq.~(\ref{eq:mu}), as a function of $\beta$ for fixed $\Om$ and $\Ok$ and three values of $c$.
Each curve ends where $L_0$ no longer exists (the minimum of $\rho_d(L)$ becomes larger than
today's density). Middle and right: future equation of state and dark-energy density for the
unrestricted best fit (inconsistent, $\mu=+0.30$), the self-consistent best fit and HDE.}
\label{fig:consistency}
\end{figure*}

The condition of Sect.~\ref{sec:consistency} has a large effect. Figure~\ref{fig:consistency}
(left) shows the mismatch $\mu$ as a function of $\beta$ for fixed $\Om$ and $\Ok$. It is zero
up to a threshold, which decreases with $c$, and then it increases quickly. At the
unrestricted best fit, it is more than $30\%$. In the unrestricted posteriors, the fraction of
self-consistent samples is $f_c=0.49$ for the full model and $0.42$ for the truncated model
(curved), and $0.50$ and $0.44$ in the flat case. With a larger tolerance of $1\%$, these
fractions increase by about $0.05$. For HDE, $f_c=1$. The middle and right panels show the
reason. The unrestricted best fit of the full model reaches $k_{\rm eff}=2$ at $a\simeq2.9$,
where $\wde$ crosses $-1$. After this point, its density increases very fast, by a factor of
about $10^6$ up to $a=20$. Over the whole posterior, the inconsistent samples reach a median
minimum $\wde\simeq-5$ before $a=20$, while the consistent samples reach only $\simeq-1.1$. The
self-consistent best fit tends to $\wde\simeq-0.99$, with a slowly decreasing density. These
two types of behaviour correspond to the two sides of the de Sitter repeller of
Sect.~\ref{sec:consistency}.

The condition changes the fit quality very little. The minimum $\chi^2$ increases from
$1485.44$ to $1485.65$ (full) and from $1485.43$ to $1485.66$ (truncated). It removes about
half of the prior-weighted volume, which reduces $\ln B$ by $0.7$--$0.9$. In the rest of the
paper, all results are for the self-consistent models, unless stated otherwise.

\subsection{Constraints on the main model}

\begin{figure}[t]
\centering
\includegraphics[width=0.85\textwidth]{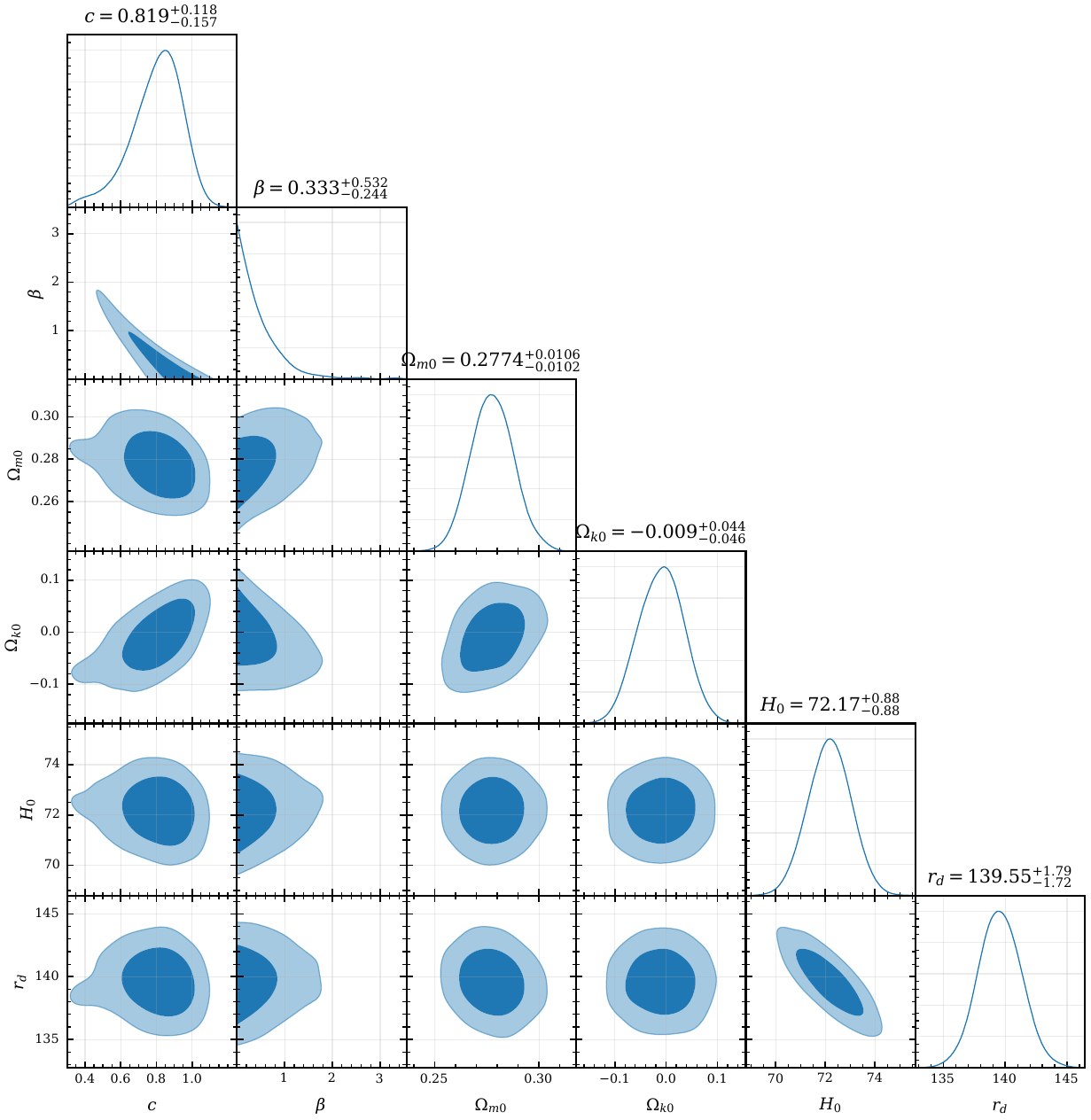}
\caption{Joint posterior of the full NTHDE model with the future-event-horizon cutoff in a
curved universe (self-consistent region), with $68\%$ and $95\%$ credible regions. The numbers
above the panels are medians with $68\%$ intervals.}
\label{fig:joint}
\end{figure}

Table~\ref{tab:post} gives the posteriors, and Fig.~\ref{fig:joint} shows the joint posterior
of the full NTHDE model with the event horizon in a curved universe. The posteriors of the
matter density, $\Om=0.2774^{+0.0106}_{-0.0102}$, the Hubble constant, $H_0=72.17\pm0.88$,
and the sound horizon, $r_d=139.6^{+1.8}_{-1.7}$ Mpc, are nearly Gaussian. The curvature is
consistent with zero, $\Ok=-0.009^{+0.044}_{-0.046}$ ($-0.094<\Ok<0.076$ at $95\%$). The
holographic parameter is $c=0.819^{+0.118}_{-0.157}$ ($0.46<c<1.04$ at $95\%$). The
deformation parameter has an upper limit but is not detected: $\beta=0.33^{+0.53}_{-0.24}$,
with $\beta<1.44$ at $95\%$. The strongest degeneracy is between $H_0$ and $r_d$, because the
BAO data measure their product. For the derived quantities, the present equation of state is
$\wde(0)=-0.967^{+0.048}_{-0.045}$, with $22\%$ of the posterior below $-1$. The deceleration
parameter is $q_0=-0.533^{+0.035}_{-0.037}$, and the transition redshift is
$z_t=0.651^{+0.023}_{-0.022}$.

\begin{table*}[t]
\centering
\caption{Posterior medians with $68\%$ intervals for the event-horizon models (NTHDE
restricted to the self-consistent region) and for flat $\Lambda$CDM. For NTHDE, the second row
gives $\beta$ (full) or $\delta$ (truncated). The last three rows are derived quantities.}
\label{tab:post}
\footnotesize\setlength{\tabcolsep}{4pt}
\begin{tabular}{lcccccc}
\toprule
Parameter & Full, curved & Full, flat & Trunc., curved & Trunc., flat & HDE, curved & $\Lambda$CDM\\
\midrule
$c$ & $0.819^{+0.118}_{-0.157}$ & $0.826^{+0.105}_{-0.138}$ & $0.848^{+0.119}_{-0.168}$ & $0.853^{+0.097}_{-0.140}$ & $0.972^{+0.087}_{-0.085}$ & --\\
$\beta$ or $\delta$ & $0.333^{+0.532}_{-0.244}$ & $0.339^{+0.490}_{-0.251}$ & $0.245^{+0.258}_{-0.176}$ & $0.250^{+0.242}_{-0.170}$ & -- & --\\
$\Omega_{m0}$ & $0.2774^{+0.0106}_{-0.0102}$ & $0.2793^{+0.0110}_{-0.0100}$ & $0.2767^{+0.0111}_{-0.0102}$ & $0.2779^{+0.0109}_{-0.0102}$ & $0.2714^{+0.0094}_{-0.0092}$ & $0.3011^{+0.0077}_{-0.0075}$\\
$\Omega_{k0}$ & $-0.009^{+0.044}_{-0.046}$ & -- & $-0.005\pm0.043$ & -- & $0.015\pm0.045$ & --\\
$H_0$ & $72.17\pm0.88$ & $72.22^{+0.88}_{-0.86}$ & $72.12^{+0.87}_{-0.86}$ & $72.16^{+0.85}_{-0.87}$ & $72.04^{+0.88}_{-0.87}$ & $72.35^{+0.85}_{-0.84}$\\
$r_d$ [Mpc] & $139.55^{+1.79}_{-1.72}$ & $139.57^{+1.78}_{-1.83}$ & $139.58^{+1.80}_{-1.74}$ & $139.58^{+1.73}_{-1.78}$ & $139.27\pm1.75$ & $139.94^{+1.75}_{-1.71}$\\
\midrule
$w_d(0)$ & $-0.967^{+0.048}_{-0.045}$ & $-0.964^{+0.039}_{-0.036}$ & $-0.958^{+0.048}_{-0.049}$ & $-0.955^{+0.037}_{-0.039}$ & $-0.919^{+0.050}_{-0.060}$ & $-1$\\
$q_0$ & $-0.533^{+0.035}_{-0.037}$ & $-0.540^{+0.036}_{-0.035}$ & $-0.531^{+0.039}_{-0.038}$ & $-0.532^{+0.035}_{-0.034}$ & $-0.518^{+0.045}_{-0.044}$ & $-0.548\pm0.011$\\
$z_t$ & $0.651^{+0.023}_{-0.022}$ & $0.654^{+0.026}_{-0.022}$ & $0.650^{+0.027}_{-0.024}$ & $0.655^{+0.023}_{-0.021}$ & $0.655^{+0.030}_{-0.024}$ & $0.668\pm0.020$\\
\bottomrule
\end{tabular}

\end{table*}

\subsection{Full versus truncated density}
\label{sec:fullvstrunc}

\begin{figure}[t]
\centering
\includegraphics[width=0.85\textwidth]{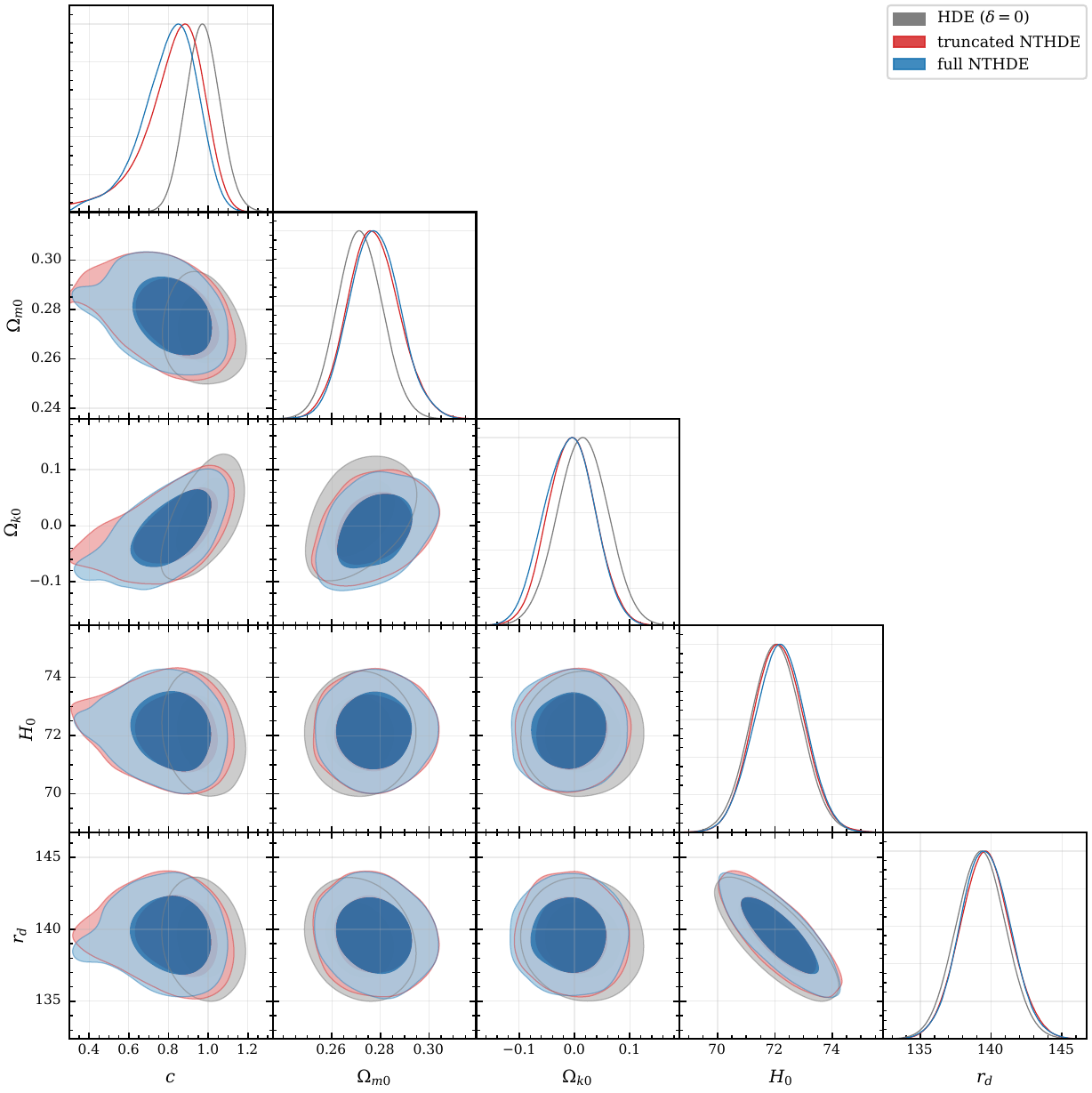}
\caption{Full NTHDE, truncated NTHDE (both self-consistent) and HDE with the event horizon in a
curved universe, for their common parameters.}
\label{fig:fullvstrunc}
\end{figure}

Figure~\ref{fig:fullvstrunc} compares the full and truncated models with HDE for their common
parameters. The two NTHDE densities give almost the same constraints and fit the data equally
well ($\chi^2_{\min}=1485.65$ and $1485.66$). However, the truncation is not a controlled
approximation. In the self-consistent posterior of the full model,
$u_0=0.39^{+0.54}_{-0.29}$ ($u_0<1.26$ at $95\%$). So, today $k_{\rm eff}=1.21^{+0.33}_{-0.16}$,
and $18^{+21}_{-13}\%$ of $\rho_d$ comes from terms beyond the Bekenstein term. If the entropy
is replaced by its second-order series, $\rho_{d0}$ changes by up to $6.6\%$ ($95\%$). In the
truncated model, the constant term alone gives $11^{+13}_{-8}\%$ of the present dark-energy
density, and it acts like a cosmological constant. Therefore, the two fits are equally good
because the three-term form is flexible, not because the series is accurate.

\subsection{Event versus apparent horizon}

With the apparent horizon, NTHDE is disfavoured by the data. The best fits are worse than
$\Lambda$CDM by $\Delta\chi^2=+4.7$ (full, curved) to $+15.4$ (truncated, flat). The
posteriors also collect near the lower limit of the matter-density prior:
$\Om=0.115^{+0.020}_{-0.011}$ for the full curved model and $0.110^{+0.015}_{-0.007}$ for the
flat one, with derived values $c=0.776^{+0.025}_{-0.049}$ and $\wde(0)=-0.84^{+0.02}_{-0.04}$
(curved). Figure~\ref{fig:bands} shows the reason. With $L=R_A$, the dark energy behaves like
pressureless matter at high redshift ($\wde\to0$), as already noted for the original model
\cite{Moradpour:2020dfm}, and it takes over part of the matter density. When the lower limit
of the prior is reduced to the baryon density, $\Om\ge0.04$ (Sect.~\ref{sec:robust}), the
posteriors move to $\Om=0.054$--$0.072$. So the dark energy takes the role of dark matter, and
even then the evidence strongly disfavours the model, with $\ln B=-7.5$ to $-9.6$. The apparent
horizon does not depend on the future evolution, so it has no self-consistency problem of the
type discussed in Sect.~\ref{sec:consistency}. However, it can describe these data only if the
dark energy replaces dark matter. This disagrees with the independent evidence for clustering
dark matter from the CMB acoustic peaks \cite{Planck:2018vyg} and from gravitational lensing
of merging clusters \cite{Clowe:2006eq}.

\subsection{Curvature and the individual data sets}

\begin{figure}[t]
\centering
\includegraphics[width=0.85\textwidth]{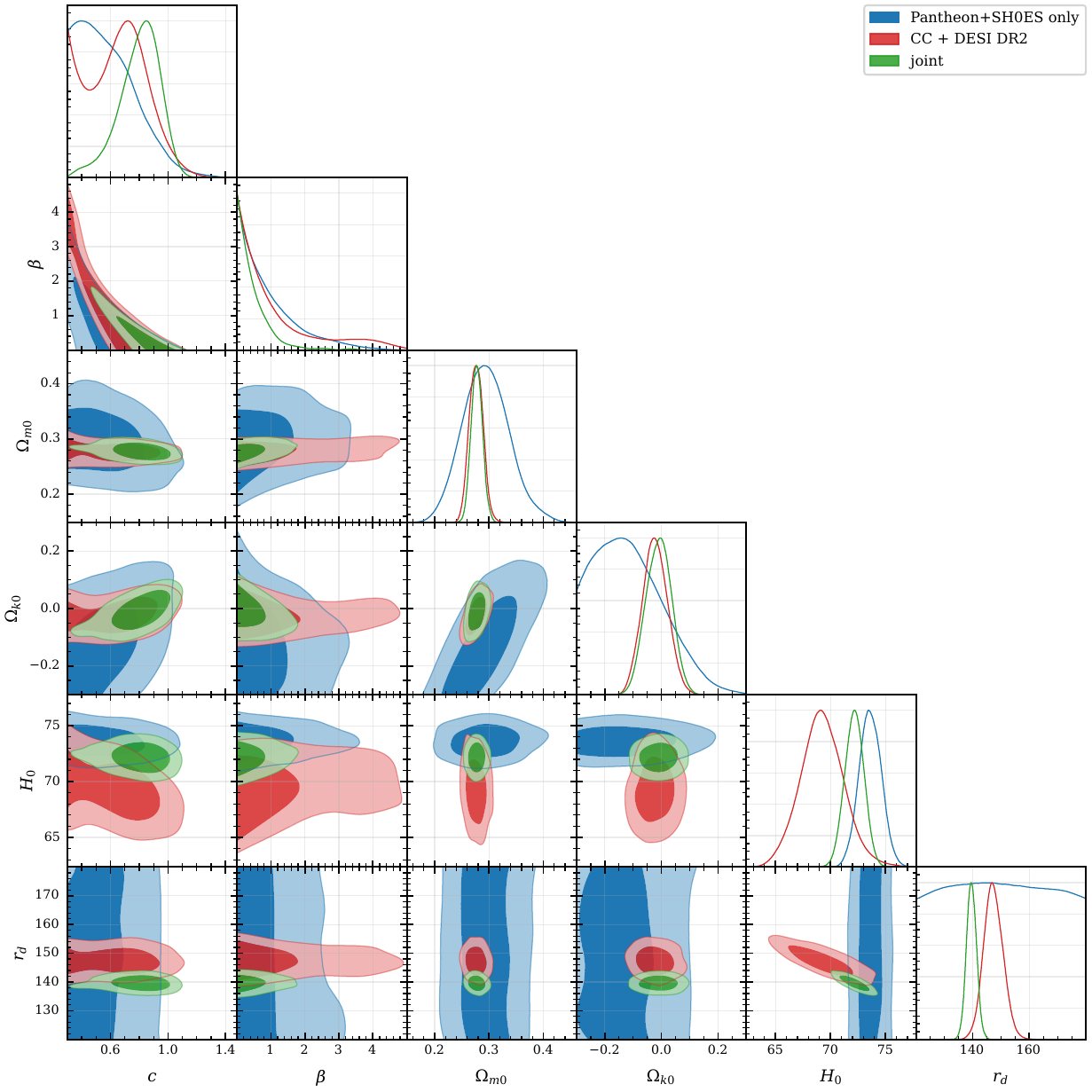}
\caption{Full NTHDE with the event horizon (self-consistent): supernovae alone, CC + DESI DR2
alone, and their combination. The CC + DESI DR2 chain reaches $43$ autocorrelation times for
$\beta$, which these data hardly constrain, and more than $59$ for the other parameters.}
\label{fig:probes}
\end{figure}

Figure~\ref{fig:probes} compares the different data sets for the main model. Supernovae alone
give $H_0=73.60^{+1.04}_{-0.98}$ through the Cepheid calibration, but they constrain the
curvature only to $\Ok=-0.12^{+0.13}_{-0.12}$ and give no information on $r_d$. Chronometers
and DESI DR2 constrain the curvature better, $\Ok=-0.024^{+0.046}_{-0.042}$, and give
$r_d=147.2^{+3.5}_{-3.2}$ Mpc. However, their value $H_0=69.16^{+2.00}_{-1.83}$ is weakly
constrained because of the $H_0$--$r_d$ degeneracy. The two independent values of $H_0$
differ by $2.0\sigma$. When all parameters are considered, combining the two data sets gives
$\Delta Q=8.6$ for $2.4$ effectively shared parameters. This is a $2.3\sigma$ difference
($p=0.02$). Thus, the two data sets are in mild tension but are not incompatible, and it is
reasonable to combine them. The joint fit lies between them, and the shift of $r_d$ to
$139.6$ Mpc is the same tension seen in the sound horizon. In every event-horizon model, the
curvature is consistent with zero, and the flat and curved versions have almost the same
$\chi^2_{\min}$ (Table~\ref{tab:comparison}). Hence, our conclusions do not depend on the
assumption of a flat universe.

\subsection{Model comparison}

Table~\ref{tab:comparison} gives the model comparison. Compared with flat $\Lambda$CDM, the
event-horizon models improve the fit by at most $\Delta\chi^2=-1.96$. HDE alone already gives
$-1.56$, so the Tsallis deformation adds only $-0.3$ to $-0.4$. The information criteria favour
$\Lambda$CDM: for the event-horizon NTHDE models, $\Delta\mathrm{AIC}=+2.1$ to $+4.1$,
$\Delta\mathrm{BIC}=+13$ to $+20$ and $\Delta\mathrm{DIC}=+1.1$ to $+2.9$. The Bayesian
evidence gives $\ln B=-2.5$ to $-5.4$. This is moderate evidence for $\Lambda$CDM, and it
becomes strong for the full curved model. The penalty increases with the number of parameters
and with the prior volume of the deformation parameter. The apparent-horizon models are
strongly disfavoured, with $\ln B=-9.8$ to $-12.4$ and $\Delta\mathrm{AIC}\ge+8.7$. Curved
$\Lambda$CDM is weakly disfavoured compared with the flat case, $\ln B=-1.3$.

The Bayes factors depend on the prior ranges. When the range of the deformation parameter is
halved or doubled, $\ln B$ changes by $\pm0.7$: from $-5.4$ to $-4.7$ ($\beta<2.5$) and
$-6.1$ ($\beta<10$) for the full curved model, and from $-2.5$ to $-1.9$ and $-3.2$ for the
truncated flat model. Therefore, the preference for $\Lambda$CDM is not caused by the choice of
priors. If the prior is normalised only over the allowed region, $\ln B$ of the event-horizon
NTHDE models increases by $1.5$--$2.2$, to between $-1.0$ and $-3.3$. For HDE it becomes
$-2.9$, and for the apparent-horizon models it lies between $-8.6$ and $-10.8$. With both
choices, the evidence does not favour NTHDE over $\Lambda$CDM.

\begin{table*}[t]
\centering
\caption{Model comparison with respect to flat $\Lambda$CDM ($\chi^2_{\min}=1487.61$). EH and
AH denote the event-horizon and apparent-horizon cutoffs, and $k$ is the number of free
parameters (the supernova magnitude, common to all models, is not counted). For the EH NTHDE
models, the self-consistency condition is imposed, and the errors of $\ln B$ include the
binomial error of the self-consistent fraction $f_c$. $\ln B$ is given for a prior that is
uniform over the full parameter box; the other normalisation is discussed in the text.}
\label{tab:comparison}
\footnotesize\setlength{\tabcolsep}{5pt}
\begin{tabular}{llccccccc}
\toprule
Model & Cutoff, geometry & $k$ & $\chi^2_{\min}$ & $\Delta\chi^2$ & $\Delta$AIC & $\Delta$BIC & $\Delta$DIC & $\ln B$\\
\midrule
$\Lambda$CDM & flat & 3 & 1487.61 & $0$ & $0$ & $0$ & $0$ & $0$\\
$\Lambda$CDM & curved & 4 & 1486.96 & $-0.65$ & $+1.35$ & $+6.79$ & $+2.17$ & $-1.33\pm0.23$\\
HDE & EH, curved & 5 & 1486.05 & $-1.56$ & $+2.44$ & $+13.32$ & $+2.32$ & $-2.98\pm0.24$\\
\midrule
full NTHDE & EH, curved & 6 & 1485.65 & $-1.96$ & $+4.04$ & $+20.36$ & $+2.88$ & $-5.42\pm0.25$\\
full NTHDE & EH, flat & 5 & 1485.68 & $-1.93$ & $+2.07$ & $+12.95$ & $+1.66$ & $-4.02\pm0.25$\\
trunc.\ NTHDE & EH, curved & 6 & 1485.66 & $-1.95$ & $+4.05$ & $+20.37$ & $+2.36$ & $-4.18\pm0.25$\\
trunc.\ NTHDE & EH, flat & 5 & 1485.70 & $-1.91$ & $+2.09$ & $+12.97$ & $+1.09$ & $-2.53\pm0.24$\\
\midrule
full NTHDE & AH, curved & 5 & 1492.29 & $+4.68$ & $+8.68$ & $+19.56$ & $+11.20$ & $-9.80\pm0.26$\\
full NTHDE & AH, flat & 4 & 1495.91 & $+8.30$ & $+10.30$ & $+15.74$ & $+13.13$ & $-10.39\pm0.26$\\
trunc.\ NTHDE & AH, curved & 5 & 1495.83 & $+8.22$ & $+12.22$ & $+23.10$ & $+14.28$ & $-10.20\pm0.25$\\
trunc.\ NTHDE & AH, flat & 4 & 1503.01 & $+15.40$ & $+17.40$ & $+22.84$ & $+21.55$ & $-12.40\pm0.25$\\
\bottomrule
\end{tabular}

\end{table*}

\subsection{Results of the additional tests}
\label{sec:robust}

Table~\ref{tab:robust} gives the results of the four additional tests. Without the Cepheid
calibration, the expansion rate is fixed by the chronometers and BAO, and we get $H_0\simeq68$
and $r_d\simeq147$ Mpc in every model. With these supernova samples, the holographic models fit
better than $\Lambda$CDM by a larger amount: $\Delta\chi^2\simeq-4.7$ to $-4.9$ with Union3 and
$-3.6$ to $-3.9$ with DES-Dovekie. This agrees with the preference for evolving dark energy
that these samples show when they are combined with DESI DR2 \cite{DESI:2025zgx,Wang:2026dde}.
For two or three extra parameters, these improvements correspond to $1.1$--$1.7\sigma$.
However, HDE already gives the same improvement, and the Tsallis deformation adds at most
$\Delta\chi^2=-0.4$. Therefore, the evidence still favours $\Lambda$CDM over NTHDE, with
$\ln B=-2.7$ to $-4.3$ (Union3) and $-3.3$ to $-4.7$ (DES-Dovekie). For HDE, it is
inconclusive with Union3 ($-0.9$) and weakly against HDE with DES-Dovekie ($-1.9$). If the
prior is normalised over the allowed region, the NTHDE values increase to between $-0.6$ and
$-2.7$. The evidence then becomes inconclusive in some cases, but it never favours NTHDE. In
these runs without Cepheid calibration, the lower tail of $c$ reaches the prior limit
$c=0.3$, so the intervals on $c$ are partly limited by the prior. The self-consistency
condition removes a large part of the NTHDE posterior in every case, $f_c=0.22$--$0.43$. The
correlated chronometer systematic changes the main result only a little ($\Delta\chi^2=-2.8$,
$\ln B=-4.9$ instead of $-5.4$). With the baryon-density prior, the apparent-horizon models are
still strongly disfavoured. Hence, our conclusions do not depend on the supernova sample, the
chronometer covariance or the matter-density prior.

\begin{table*}[!htbp]
\centering
\caption{Additional tests. For each data combination, $\Delta\chi^2$, $\Delta$AIC, $\Delta$BIC
and $\ln B$ are given with respect to flat $\Lambda$CDM with the same data. $f_c$ is the
self-consistent fraction of the event-horizon NTHDE posteriors, and their $\chi^2_{\min}$ and
$\ln B$ include the condition. $H_0$ is in km s$^{-1}$ Mpc$^{-1}$, and the posteriors are
medians with $68\%$ intervals from nested sampling.}
\label{tab:robust}
\footnotesize\setlength{\tabcolsep}{3.5pt}
\begin{tabular}{lccccccccc}
\toprule
Model & $k$ & $\chi^2_{\min}$ & $\Delta\chi^2$ & $\Delta$AIC & $\Delta$BIC & $\ln B$ & $f_c$ & $H_0$ & $\Omega_{m0}$\\
\midrule
\multicolumn{10}{l}{\emph{Union3 + CC + DESI DR2 (no Cepheid calibration)}}\\
$\Lambda$CDM, flat & 3 & 53.62 & -- & -- & -- & -- & -- & $68.7\pm1.7$ & $0.304\pm0.008$\\
$\Lambda$CDM, curved & 4 & 51.99 & $-1.63$ & $+0.37$ & $+2.57$ & $-0.82$ & -- & $68.1\pm1.7$ & $0.294\pm0.012$\\
HDE, EH, curved & 5 & 48.89 & $-4.73$ & $-0.73$ & $+3.68$ & $-0.86$ & -- & $67.0\pm1.7$ & $0.271^{+0.010}_{-0.009}$\\
full NTHDE, EH, curved & 6 & 48.90 & $-4.72$ & $+1.28$ & $+7.89$ & $-4.30$ & 0.28 & $67.9^{+1.7}_{-1.9}$ & $0.279^{+0.013}_{-0.012}$\\
full NTHDE, EH, flat & 5 & 48.80 & $-4.82$ & $-0.82$ & $+3.59$ & $-2.71$ & 0.26 & $68.4^{+1.9}_{-2.1}$ & $0.284^{+0.014}_{-0.011}$\\
trunc.\ NTHDE, EH, curved & 6 & 48.76 & $-4.86$ & $+1.14$ & $+7.76$ & $-2.66$ & 0.22 & $67.7^{+1.7}_{-1.8}$ & $0.277^{+0.011}_{-0.009}$\\
\midrule
\multicolumn{10}{l}{\emph{DES-Dovekie + CC + DESI DR2 (no Cepheid calibration)}}\\
$\Lambda$CDM, flat & 3 & 1660.09 & -- & -- & -- & -- & -- & $68.6\pm1.6$ & $0.306^{+0.008}_{-0.007}$\\
HDE, EH, curved & 5 & 1656.54 & $-3.55$ & $+0.45$ & $+11.51$ & $-1.89$ & -- & $67.8^{+1.7}_{-1.5}$ & $0.271\pm0.009$\\
full NTHDE, EH, curved & 6 & 1656.17 & $-3.92$ & $+2.08$ & $+18.67$ & $-4.69$ & 0.43 & $67.9^{+1.6}_{-1.4}$ & $0.281^{+0.011}_{-0.013}$\\
trunc.\ NTHDE, EH, curved & 6 & 1656.27 & $-3.82$ & $+2.18$ & $+18.77$ & $-3.33$ & 0.32 & $67.9^{+1.9}_{-1.5}$ & $0.276^{+0.010}_{-0.011}$\\
\midrule
\multicolumn{10}{l}{\emph{Pantheon+SH0ES + CC with a $5\%$ correlated systematic + DESI DR2}}\\
$\Lambda$CDM, flat & 3 & 1483.75 & -- & -- & -- & -- & -- & $73.2\pm1.0$ & $0.303\pm0.008$\\
full NTHDE, EH, curved & 6 & 1480.93 & $-2.82$ & $+3.18$ & $+19.50$ & $-4.88$ & 0.43 & $73.0\pm0.9$ & $0.277^{+0.011}_{-0.010}$\\
\midrule
\multicolumn{10}{l}{\emph{Pantheon+SH0ES + CC + DESI DR2, prior $\Omega_{m0}\ge0.04$}}\\
$\Lambda$CDM, flat & 3 & 1487.61 & -- & -- & -- & -- & -- & $72.3^{+0.9}_{-0.8}$ & $0.301\pm0.008$\\
full NTHDE, AH, curved & 5 & 1488.48 & $+0.87$ & $+4.87$ & $+15.75$ & $-8.09$ & -- & $72.3^{+0.8}_{-0.9}$ & $0.072^{+0.034}_{-0.022}$\\
full NTHDE, AH, flat & 4 & 1489.92 & $+2.31$ & $+4.31$ & $+9.75$ & $-7.53$ & -- & $72.6\pm0.9$ & $0.059^{+0.024}_{-0.014}$\\
trunc.\ NTHDE, AH, curved & 5 & 1491.61 & $+4.00$ & $+8.00$ & $+18.88$ & $-8.66$ & -- & $72.4\pm0.8$ & $0.067^{+0.061}_{-0.020}$\\
trunc.\ NTHDE, AH, flat & 4 & 1495.18 & $+7.57$ & $+9.57$ & $+15.01$ & $-9.57$ & -- & $72.7^{+0.9}_{-0.8}$ & $0.054^{+0.024}_{-0.011}$\\
\bottomrule
\end{tabular}

\end{table*}
\FloatBarrier

\subsection{Background evolution and the future}

\begin{figure*}[t]
\centering
\includegraphics[width=\textwidth]{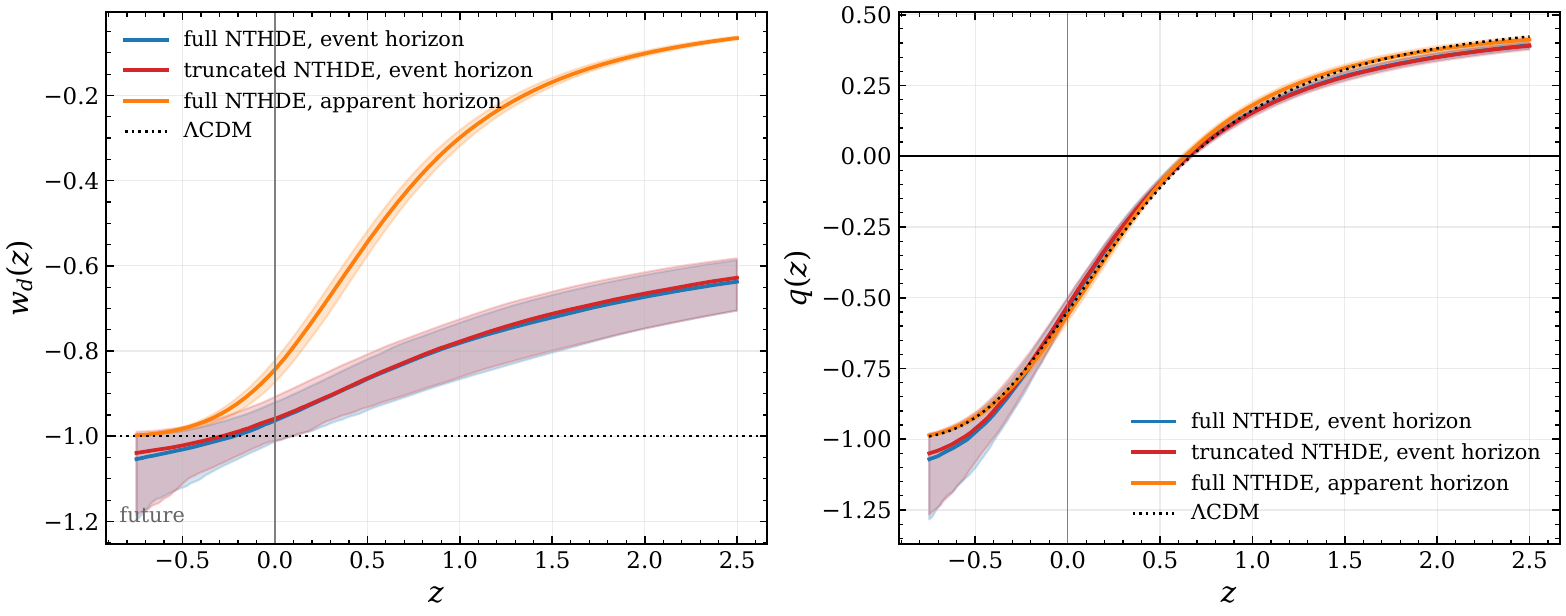}
\caption{Equation of state (left) and deceleration parameter (right) with $68\%$ bands for full
and truncated NTHDE with the event horizon (self-consistent) and full NTHDE with the apparent
horizon, in curved universes. The region $z<0$ is the future, up to $a=4$.}
\label{fig:bands}
\end{figure*}

Figure~\ref{fig:bands} shows $\wde(z)$ and $q(z)$ with $68\%$ bands, including the future
($z<0$). Today, the two event-horizon NTHDE models behave like quintessence,
$\wde(0)\simeq-0.96$. Their $q(z)$ is almost the same as in $\Lambda$CDM over the redshift
range of the data, with $z_t\simeq0.65$. For most of the self-consistent posterior, $\wde$
crosses $-1$ slowly in the near future (the median reaches $\wde\simeq-1.05$ at $a=4$). In
every self-consistent sample that crosses $-1$ before $a=20$, the crossing happens where the
horizon stops growing ($L'=0$), with $k_{\rm eff}<2$. This is the same mechanism as in HDE
with $c<1$, and it gives a mildly phantom future, as in HDE. The rapid crossing at
$k_{\rm eff}=2$ found in Sect.~\ref{sec:consistency} happens only in the excluded region.

\section{Discussion}
\label{sec:discussion}

\emph{Why NTHDE fits slightly better.} Compared with $\Lambda$CDM, the event-horizon models
improve the fit by at most $|\Delta\chi^2|\simeq2$, and HDE already gives most of this
improvement. In the truncated density, the constant term acts as a cosmological constant. In
the full density, the part of $(e^u-1)/u$ above unity plays the same role, and it changes
slowly while $u<1$. The data allow this extra freedom but do not need it. Therefore, every
criterion that penalises extra parameters or prior volume prefers $\Lambda$CDM.

\emph{The truncated density.} Since the expansion parameter reaches $u=O(1)$, the three-term
density should not be taken as an approximation of NTHDE for the parameter values allowed by
the data. It is still a valid phenomenological model, and its constraints are close to those
of the full model. However, results from the truncated model should not be assigned to the
Tsallis entropy of Eq.~(\ref{eq:ST}) unless they are checked with the full density.

\emph{Thermodynamic interpretation.} For any holographic density $\rho_d\propto S(A)/A^2$,
the density is stationary, and $\wde$ crosses $-1$ along a solution with $L'\neq0$, exactly
where the effective entropy exponent is $k_{\rm eff}=2$. The same value appears, in a different
framework, in the generalised-second-law analysis of Ref.~\cite{Cruz:2026gsl}. That work
connects a smooth crossing of the phantom divide with a quadratic scaling of entropy with area.
It studies entropic changes of the Friedmann equations and not holographic dark energy, so this
agreement is only an indication and not a derivation. For the exponential entropy of
Eq.~(\ref{eq:ST}), $k_{\rm eff}$ grows without limit at large area, as for the Kaniadakis
entropy discussed in that work.

\emph{Comparison with other studies.} Non-flat HDE and interacting HDE were recently tested
with DESI DR2 combined with DES supernovae, CMB distance priors and a local $H_0$ prior. In
that study, the interacting models preferred a closed universe, and all holographic models
were disfavoured compared with $\Lambda$CDM \cite{Shen:2026ihde}. Without CMB information, we
find a curvature consistent with zero and the same general preference for $\Lambda$CDM. The
choice of supernova sample is important for dynamical dark energy: Pantheon+, Union3 and DES
give different significances for evolving dark energy with DESI DR2 \cite{Wang:2026dde}. We
find the same trend, with a larger improvement in fit for the samples without Cepheid
calibration. However, in the holographic models this improvement comes from the HDE term and
not from the Tsallis deformation. Barrow and Tsallis holographic dark energy, which have a
power-law deformation of the area law, are also consistent with DESI DR2, supernova and
chronometer data. They are not favoured over $\Lambda$CDM by information criteria, and they do
not reduce the $H_0$ tension \cite{Luciano:2025bts}. Barrow holographic dark energy coupled to
non-cold dark matter has also been tested with similar data \cite{AlMamon:2026bhde}. Our
results for the exponential Tsallis entropy show the same pattern. With the Granda--Oliveros
cutoff, the same data also give a curvature consistent with zero, and the evidence also favours
$\Lambda$CDM, except with Union3, where the two models are on a par \cite{Sharma:2026goNTHDE}.
Our results with the future event horizon agree with this picture.

\emph{What is new in the consistency test.} It has been known for a long time that the
event-horizon cutoff depends on the future evolution \cite{Li:2004rb,Cai:2007us}, and this led
to other cutoffs \cite{Cai:2007us,Gao:2007ep}. However, as far as we know, the condition that
the horizon obtained from today's density must be equal to the true future event horizon of the
model has not been tested in observational studies of generalised-entropy holographic models.
In these models the condition is important, because the density is not monotonic in $L$.

\emph{Limitations.} Our analysis uses only the background expansion. Perturbations and
growth-rate data, which could separate the models better, are not included. We have changed
the supernova sample, the chronometer systematics and the matter-density prior
(Sect.~\ref{sec:robust}). However, the chronometer test uses a simple correlated error instead
of the full covariance of Ref.~\cite{Moresco:2020fbm}, and we did not use CMB information on
$r_d$. The self-consistency condition is imposed on the posterior samples. If its tolerance is
increased to $1\%$, the kept fractions change by about $0.05$. Finally, the Bayes factors
depend on the prior widths, as discussed above.

\section{Conclusions}
\label{sec:conclusions}

In this work, we have tested new Tsallis holographic dark energy with Pantheon+SH0ES, cosmic
chronometers and DESI DR2 data. We compared the full exponential density with its truncated
three-term form, the event horizon with the apparent horizon, and flat with curved universes,
keeping the sound horizon and the supernova magnitude free. Our main results are as follows.

\begin{enumerate}
\item The future-event-horizon cutoff has a self-consistency condition. The horizon obtained
from today's dark-energy density must be equal to the event horizon of the expansion predicted
by the same model. This condition always holds in standard HDE, but it fails for about half of
the NTHDE posterior. The reason is a de Sitter fixed point of the late-time dynamics, which
acts as a repeller, and the solutions that cross it grow without control. Imposing the
condition changes the fit by only $\Delta\chi^2\simeq+0.2$.
\item With the condition imposed, the full NTHDE model gives $\Om=0.2774^{+0.0106}_{-0.0102}$,
$\Ok=-0.009^{+0.044}_{-0.046}$, $H_0=72.17\pm0.88$, $r_d=139.6^{+1.8}_{-1.7}$ Mpc,
$c=0.82^{+0.12}_{-0.16}$ and the deformation parameter $\beta<1.44$ at $95\%$. The present
equation of state is $\wde(0)=-0.967^{+0.048}_{-0.045}$, and the transition redshift is
$z_t=0.65\pm0.02$.
\item The full and truncated densities fit the data equally well, but the truncation is not a
controlled approximation. The expansion parameter reaches $u=O(1)$, so the three-term form
should be treated as a separate phenomenological model.
\item NTHDE improves the fit over flat $\Lambda$CDM by at most $\Delta\chi^2\simeq-2$, and most
of this improvement is already given by standard HDE. AIC, BIC and DIC favour $\Lambda$CDM.
The Bayesian evidence never favours NTHDE: $\ln B=-2.5$ to $-5.4$, or $-1.0$ to $-3.3$ if the
prior is normalised over the region where the model has self-consistent solutions. These values
change little when the prior range of the deformation parameter is halved or doubled.
\item These conclusions also hold with the Union3 and DES-Dovekie supernova samples, which have
no Cepheid calibration, and with a correlated systematic error in the chronometer data. With
these two samples, the improvement in fit increases to $\Delta\chi^2\simeq-4$ (a
$1.1$--$1.7\sigma$ effect). However, standard HDE already gives this improvement, and the
evidence still never favours NTHDE.
\item With the apparent horizon, NTHDE behaves like matter at high redshift. It fits the data
only if it replaces dark matter, so that the matter density falls to the baryon value. This
model is strongly disfavoured ($\ln B\simeq-7.5$ to $-12$).
\item Chronometers and DESI DR2 alone give $H_0=69.2\pm1.9$ and $r_d=147.2^{+3.5}_{-3.2}$ Mpc.
This $H_0$ is $2.0\sigma$ lower than the supernova value, and NTHDE does not reduce this
tension. A model-independent reconstruction from the chronometers gives $q=-0.48\pm0.12$ at
$z=0.05$ (Appendix~\ref{app:ann}).
\end{enumerate}

\noindent The self-consistency test is not limited to NTHDE. It can be used for any holographic
dark-energy model with a future-event-horizon cutoff whose density is not monotonic in the
cutoff length, for example models based on the Sharma--Mittal entropy or other exponential-type
entropies. We suggest that this test should be done routinely. Since the dark energy crosses
the phantom divide where the effective entropy exponent is equal to two, the test also connects
holographic dark energy with thermodynamic selection criteria for generalised entropies.

\appendix
\section{Model-independent reconstruction of the expansion history}
\label{app:ann}

When the curvature is free, $H(z)$ cannot be reconstructed from supernovae without assuming a
value of $\Ok$, so we use the chronometers. We use a bootstrap deep ensemble
\cite{Lakshminarayanan:2016aa} of $60$ networks (one hidden layer with eight $\tanh$ units,
weight decay $0.1$). Each network maps $z\mapsto[H,\ln\sigma_{\rm int}]$ with a heteroscedastic
Gaussian likelihood, and $q=(1+z)H'/H-1$ is obtained by automatic differentiation. The networks
are small and strongly regularised, because only $32$ data points are available. The
reconstructed $H(z)$ does not depend much on these choices, but its derivatives do.

\begin{figure*}[h]
\centering
\includegraphics[width=\textwidth]{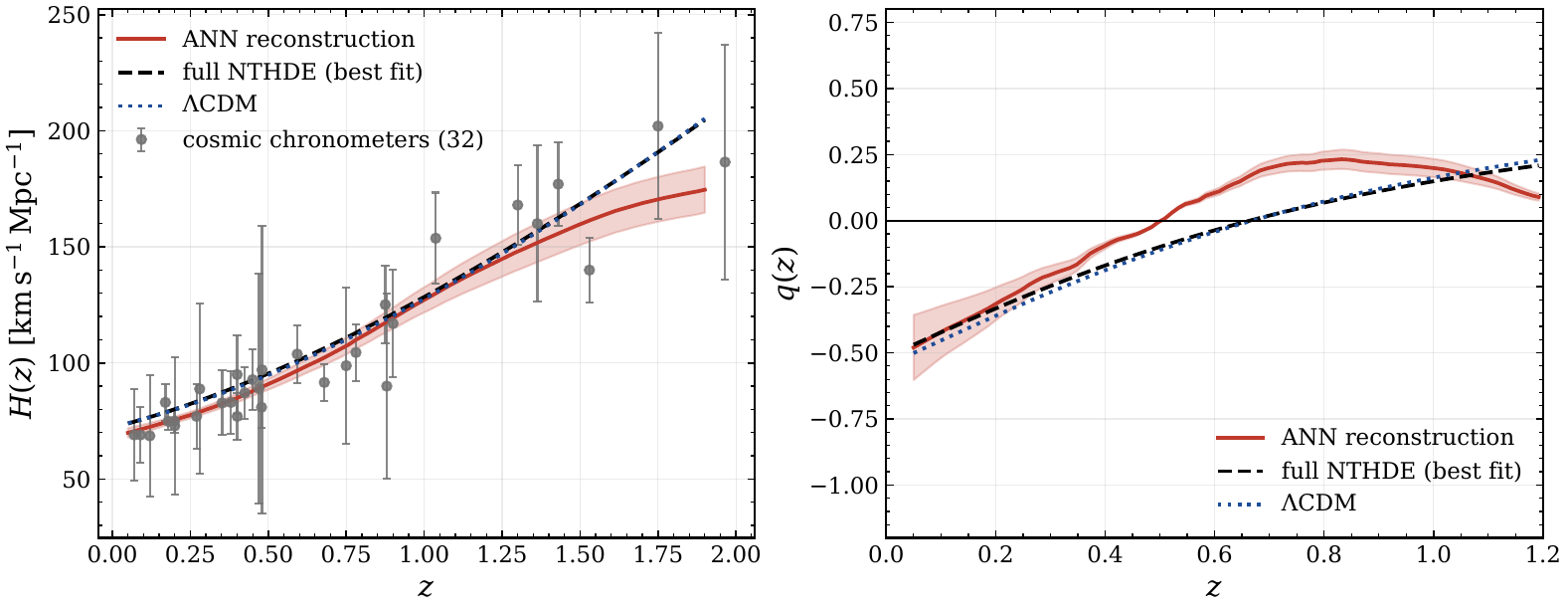}
\caption{Neural-network reconstruction from the $32$ chronometer measurements ($1\sigma$
ensemble band). Left: $H(z)$ with the data, the self-consistent best fit of the main model and
$\Lambda$CDM. Right: $q(z)$ for $z<1.2$, where the data constrain the derivative.}
\label{fig:ann}
\end{figure*}

At $z=0.05$, the lowest redshift of the reconstruction, the network ensemble
(Fig.~\ref{fig:ann}) gives $H=69.9\pm2.0$ and $q=-0.48\pm0.12$. This confirms the present
acceleration at about $4\sigma$ without assuming a cosmological model. On average over
$0.1<z<1.5$, the reconstruction agrees with the self-consistent best fit of the main model to
$3.6\%$ and with $\Lambda$CDM to $3.1\%$. So the chronometers alone cannot distinguish between
the two models. The reconstructed transition redshift changes by $\Delta z\sim0.5$ for
reasonable changes of the regularisation, so we do not quote it.

\section*{Declarations}

\noindent\textbf{Funding.} The authors did not receive support from any organisation for the
submitted work.

\noindent\textbf{Competing interests.} The authors declare no competing interests.

\noindent\textbf{Data availability.} The data used in this study are openly available:
Pantheon+SH0ES (\url{https://github.com/PantheonPlusSH0ES/DataRelease}), Union3 and
DES-Dovekie (\url{https://github.com/CobayaSampler/sn_data}), DESI DR2 baryon acoustic
oscillations (\url{https://github.com/CobayaSampler/bao_data}), and the cosmic-chronometer
measurements published in the references cited in the text.

\bibliographystyle{unsrt}
\bibliography{refPSK}

\end{document}